\documentclass[11pt,twoside]{article}  % Leave intact
\usepackage{apn3conf}

\begin{document}   % Leave intact

\title{Influence of the Interstellar Medium on the Shaping of Planetary Nebulae}
\titlemark{ISM Influence on Shaping PNe}

\author{Hans-Reinhard M\"uller\altaffilmark{1}}
\affil{Department of Physics and Astronomy, Dartmouth College,
       Hanover, NH 03755}

\author{Florian Kerber}
\affil{Space Telescope – European Co-ordinating Facility,
Garching, Germany}

\author{Thomas Rauch\altaffilmark{2} and Eva-Maria Pauli}
\affil{Dr.-Remeis-Sternwarte Bamberg, University of
Erlangen-N\"urnberg, Germany}

\altaffiltext{1}{IGPP, University of California, Riverside, CA
92521; E-mail: Hans.Mueller@dartmouth.edu}
\altaffiltext{2}{Institut f\"ur Astronomie und Astrophysik,
University of T\"ubingen, Germany}

%-----------------------------------------------------------------------
%            Contact Information
%-----------------------------------------------------------------------
% This information will not appear in the paper but will be used by
% the editors in case you need to be contacted concerning your
% submission.  Enter your name as the contact along with your email
% address.

\contact{Hans Mueller}
\email{hans.mueller@dartmouth.edu}

%-----------------------------------------------------------------------
%             Author Index Specification
%-----------------------------------------------------------------------
% Specify how each author name should appear in the author index.  The
% \paindex{ } should be used to indicate the primary author, and the
% \aindex for all other co-authors.  You MUST use the following
% syntax:
%
% SYNTAX:  \aindex{LASTNAME, F. M.}

\paindex{Mueller, H. R.}
\aindex{Kerber, F.}
\aindex{Rauch, T.}
\aindex{Pauli, E. M.}

%-----------------------------------------------------------------------
%             Author list for page header
%-----------------------------------------------------------------------
% Please supply a list of author last names for the page header. in
% one of these formats:
%
% EXAMPLES:
% \authormark{LASTNAME}
% \authormark{LASTNAME1 \& LASTNAME2}
% \authormark{LASTNAME1, LASTNAME2, ... \& LASTNAMEn}
% \authormark{LASTNAME et al.}
%
% Use the "et al." form in the case of seven or more authors, or if
% the preferred form is too long to fit in the header.

\authormark{M\"uller, Kerber, Rauch \& Pauli}

%-----------------------------------------------------------------------
%           Subject Index keywords
%-----------------------------------------------------------------------
% Enter up to 6 keywords describing your paper.  These will NOT be
% printed as part of your paper; however, they will be used to
% generate an object index and a subject index for the proceedings.
% There is no standard list,  however, individual object names are
% encouraged and one or two word descriptions of the topics (e.g.MHD,
% ionized gas) are useful.
%
% EXAMPLE:  \keywords{NGC 7027, AFGL 2688, HD 161796, binary stars,
%                      dust,  molecular gas}
%


%-----------------------------------------------------------------------
%                  Abstract
%-----------------------------------------------------------------------
% Type abstract in the space below.  Consult the User Guide and Latex
% Information file for a list of supported macros (e.g. for typesetting
% special symbols). Do not leave a blank line between \begin{abstract}
% and the start of your text.

\begin{abstract}          % Leave intact
The interaction of the ISM with the evolution of PNe is studied
for cases where there is a large relative velocity between ISM and
central star, for example for high-proper motion PNe. In such
cases the ISM wind already interacts strongly with the AGB wind,
and extensive PN asymmetries result.
\end{abstract}

%-----------------------------------------------------------------------
%                 Main Body
%-----------------------------------------------------------------------
% Place the text for the main body of the paper here.  You should use
% the \section command to label the various sections; use of
% \subsection is optional.  Significant words in section titles should
% be capitalized.  Sections and subsections will be numbered
% automatically.
%
% EXAMPLE:  \section{Introduction}
%           ...
%           \subsection{Our View of the World}
%           ...
%           \section{A New Approach}
%
% It is recommended that you look at the sample papers, sample1.tex
% and sample2.tex, for examples for formatting references, footnotes,
% figures, equations, html links, lists, and other special features.

\section{Introduction}

Stars, stellar winds, and planetary nebulae (PNe) are generally in
a different velocity state than the interstellar medium (ISM) in
which they are embedded.
The most accurately known example is that of the Sun and its local
ISM (LISM). Both the Sun and the LISM are in motion with respect
to the local standard of rest; the velocities combine to a solar
relative apex motion of 26 km/s.% with respect to the LISM.

The LISM is part of the Local Interstellar Cloud (LIC) which is
warm ($\sim 7000$ K) and dense ($\sim 0.2\mbox{ cm}^{-3}$ H$^0$,
$\sim 0.1\mbox{ cm}^{-3}$ H$^+$). The LIC is embedded in the Local
Bubble ($1.2\times 10^6$ K, $\sim 0.005\mbox{ cm}^{-3}$ H$^+$).
These are but two examples of wide-ranging possible interstellar
environments. Similarly, different PNe evolve embedded in a
variety of ISM environments. For larger ISM ram pressures, the
wind-wind interaction will naturally give rise to asymmetries in
the PNe.

PNe -- ISM interaction has already been studied extensively
(Borkowski, Sarazin, \& Soker 1990; Soker, Borkowski, \& Sarazin
1991). According to recent arguments, the ISM-caused asymmetry in
PNe does not only begin at an evolved stage when the thin PN shell
is sufficiently diluted as to match the ISM density. Rather, the
ISM already interacts with the AGB wind of the central star and
can shape the region of the circumstellar material, and the
proto-planetary nebula (Villaver, Garc\'{\i}a-Segura, \& Manchado
2002a, 2003; Villaver, Manchado, \& Garc\'{\i}a-Segura 2002b).
This contribution seeks to extend the published work to
high-velocity regimes which are probably responsible for the
structure of high-proper motion PNe such as the PN around Sh 2-68
(PN G030.6+06.2; e.g.\ Kerber et al.\ 2002). Sh 2-68 has a large
proper motion, and is off the geometric center of a 15' core
nebula. There is an extended tail, that extends over $\sim 30$'
(see figure in Kerber et al.\ 2003). Both these facts suggest an
ISM flow interacting with the stellar outflow beginning early on
in the star's evolution.
\begin{figure}
\epsscale{.60} \plotone{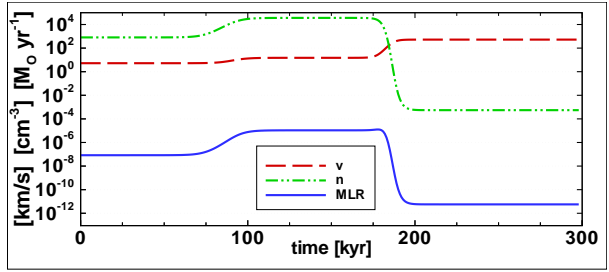} \caption{Assumed,
simplified history of the evolution of the CS.
%The initial stellar
%wind is that of a quiet AGB phase ($t < 66$ kyr); intermediate
%stage is a single, long thermal pulse (100-180 kyr), and the final
%stage is a post-AGB hot, fast white dwarf wind ($t > 200$ kyr).
The composite plot shows the wind velocity (dashed, in km/s), the
density at 0.01 pc (dot-dashed, in cm$^{-3}$), and the mass loss
rate (MLR) of the wind (solid, in $M_{\sun} \mbox{ yr}^{-1}$),
versus time in kyr.} \label{mlr-history}
\end{figure}

\section{Simple Model of Time-Dependent AGB/post-AGB Stellar Wind}

We model the wind-wind interaction from the onset of the AGB wind
to the evolved PN in an axisymmetric configuration, with the
symmetry axis containing the ISM flow vector and the CS. We use a
simplified 2D hydrodynamic numerical model (magnetic fields are
neglected) based on the Zeus3D package, with radial distance from
CS and angle from the symmetry axis as coordinates. The numerical
region is bounded by the inner (stellar wind) boundary at r = 0.01
pc (2000 AU) and the outer (ISM) boundary at r = 16 pc, and is
realized on a $430 \times 37$ grid.
The stellar wind evolution is modeled in an approximate,
simplified way. Initially, the grid is filled with a uniform ISM.
At t = 0, a slow AGB wind starts, with a mass loss rate (MLR) of
$\dot{M} = 8 \times 10^{-8} M_{\sun} \mbox{ yr}^{-1}$ and terminal
wind velocity of 7 km/s. Figure 1 shows wind velocity, wind
density at 0.01 pc, and mass loss rate in a combined plot. At
66,000 years, a transition to the thermal pulse regime starts,
reached at about 100 kyr. The AGB-TP regime is characterized by
$\dot{M} = 10^{-5} M_{\sun} \mbox{ yr}^{-1}$ and a terminal wind
velocity of 15 km/s; see Figure 1. After 80 kyr in the TP phase,
the wind transits to the post-AGB phase with $\dot{M} = 6\times
10^{-12} M_{\sun} \mbox{ yr}^{-1}$ and $v = 520$ km/s. The values
for a given CS can be expected to substantially differ from this
assumed model history; the allowed ranges for the velocity and MLR
are likely very broad.

We conduct a limited parameter survey and study six different ISM
environments, all at 8000 K: Two relative ISM velocities, 64 km/s
and 104 km/s, and three ISM hydrogen densities: 0.2, 0.1, and 0.02
cm$^{-3}$.

\section{Results}

Right from the start of the modeled evolution, the ISM interacts
with the emer\-ging AGB wind and distorts the evolution, creating
asymmetry. This is the case even for the smallest ISM density of
0.02 cm$^{-3}$. Figure 2 shows the two-dimensional density
distributions after 66 kyr of such an evolution, for all six
models. The snapshot occurred just before the gradual onset of the
TP phase. The ISM/CS wind-wind interaction is obvious. In all
cases there is an overdensity and compression in the nose
direction and a density depletion in the tail region. The
interaction manages to divert material into the post-tail region
(downwind), creating another overdensity. The different
interstellar ram pressures in the six models give rise to
different opening angles of the overdense structures, and to
different distances where these structures cross the stagnation
axis.
\begin{figure}
\epsscale{0.9} \plotone{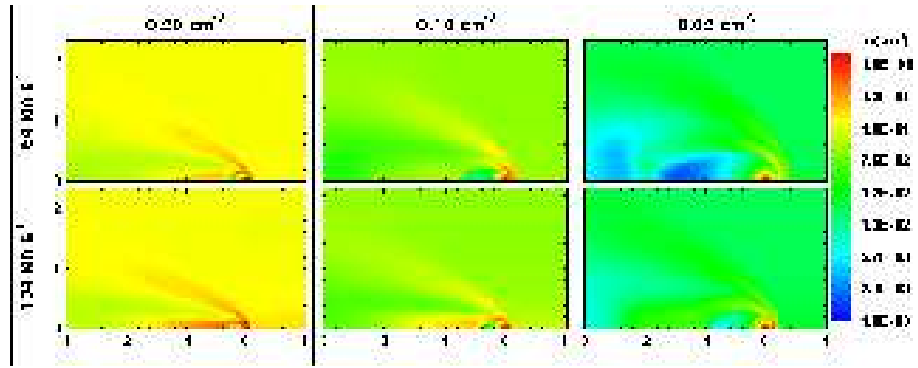}
%\epsfxsize=4.8in \epsfysize=1.9in
%\plotfiddle{mueller_2c.png}{1.7in}{0.}{70.}{70.}{-100}{0}
\caption{Snapshot, at 66 kyr, of the evolved quiet-time AGB wind
interacting with the ISM. The CS is at (0, 0), and the ISM wind
enters from the right. Color-coded number density maps are shown
for all six models. All distances in parsec.} \label{pre-tp}
\end{figure}
\begin{figure}
\epsscale{0.9} \plotone{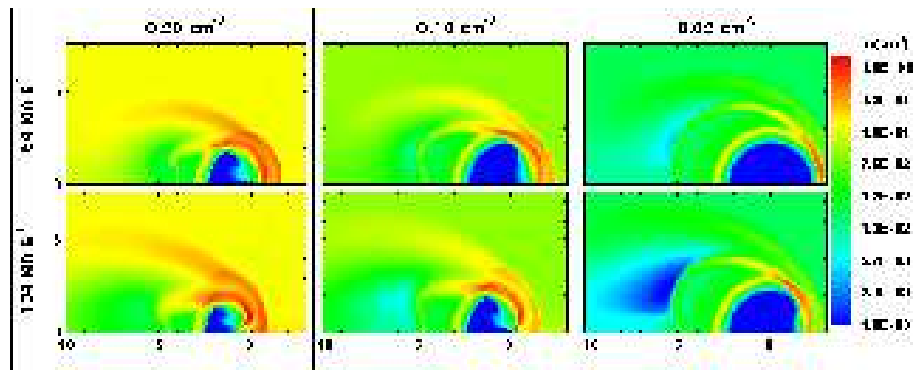}
% \epsfxsize=3.8in
%\epsfysize=1.9in
%\plotfiddle{mueller_3c.png}{1.7in}{0.}{70.}{70.}{-1000}{0}
\caption{Color maps of number density, for all six models, at 220
kyr, soon after the post-AGB wind has reached the dense
structures.} \label{post-tp}
\end{figure}

The dense stellar wind of the AGB thermal pulse cannot restore
symmetry, even when this phase is rather long as it is in our
evolution model. The increased density and velocity of the AGB-TP
wind shifts the pressure balance and lets the existing structures
grow in size. This growth has not ended when it is overtaken by
the initial 'blast' of the post-AGB wind. Figure 3 shows the
density distribution relatively soon after the white dwarf wind
has reached the dense structures, at 220 kyr. The wave compresses
the structures, and empties a large interior region. The post-AGB
wind upsets the balance through a decreased ram pressure which
allows the ISM ram pressure to diffuse some of the dense nose
material towards the CS, in some cases even drilling a hole at the
nose.

When tracked further, the interaction will separate the CS from
the PN for the larger ISM ram pressures. The general expansion of
the nebula dilutes it, and the ISM wind ablates it further.

\section{Conclusions}

If the ISM is anything but very tenuous, the sellar wind of an
AGB-branch star will interact with it. As a minimal consequence,
present even for modest ISM ram pressures, the ISM will take
emerging structures like PN shells downwind and hence the CS will
be off the geometric center of the nebula.

For the ISM parameters studied, non-spherical structures develop
already in the AGB phase. Spherical symmetry is never regained. We
have run the exact same stellar evolution (stellar wind history)
for six different ISM environments. The modest variation of ISM
ram pressure by a factor of 200 resulted in vastly different AGB
end states and PNe.

In particular for high-proper motion PNe, the wind-ISM interaction
hence needs to be taken into account. For a realistic model that
places constraints on PNe and the CS from observations, a
realistic mass loss history is needed, expanding on the toy model
presented here (Figure 1). Villaver et al.\ (2002a,b, 2003) use
many such realistic ingredients, including wind evolution derived
from stellar evolution, photoionization, radiative cooling, etc.
Even more physics is likely needed for a realistic model, such as
stellar and interstellar magnetic fields, stellar rotation, and
the ionization state of the ISM.

\acknowledgments

HRM gratefully acknowledges NASA grants NAG5-12628 to the
University of Delaware and NAG5-13611 to Dartmouth College.

%-----------------------------------------------------------------------
%                 References
%-----------------------------------------------------------------------
% List your references below within the reference environment
% (i.e. between the \begin{references} and \end{references} tags).
% Each new reference should begin with a \reference command which sets
% up the proper indentation.  Observe the following order when listing
% bibliographical information for each reference:  author name(s),
% publication year, journal name, volume, and page number for
% articles.  Note that many journal names are available as macros; see
% the User Guide listing "macro-ized" journals.
%
% EXAMPLE:  \reference Hagiwara, K., \& Zeppenfeld, D.\  1986,
%                Nucl.Phys., 274, 1
%           \reference H\'enon, M.\  1961, Ann.d'Ap., 24, 369
%           \reference King, I.\ R.\  1966, \aj, 71, 276
%           \reference King, I.\ R.\  1975, in Dynamics of Stellar
%                Systems, ed.\ A.\ Hayli (Dordrecht: Reidel), 99
%
% Note the following tricks used in the example above:
%
%   o  \& is used to format an ampersand symbol (&).
%   o  \'e puts an accent agu over the letter e.  See the User Guide
%      and the sample files for details on formatting special
%      characters.
%   o  "\ " after a period prevents LaTeX from interpreting the period
%      as an end of a sentence.
%   o  \aj is a macro that expands to "Astron. J."  See the User Guide
%      for a full list of journal macros
%


\begin{references}

\reference Borkowski, K.J., Sarazin, C.L., \& Soker, N.\ 1990,
\apj, 360, 173

\reference Kerber, F., Guglielmetti, F., Mignani, R., \& Roth, M.\
2002, \aap, 381, L9
% Proper motion of the central star of the PNe Sh 2-68

\reference Kerber, F., Guglielmetti, F., Mignani, R., \& Roth, M.\
2003, 'Sh 2-68 -- A planetary nebula leaving its mark on the
interstellar medium', in: Proceedings of IAU Symposium 209
'Planetary Nebulae", Canberra, Australia 2001, in press

\reference Soker, N., Borkowski, K.J., \& Sarazin, C.L.\ 1991,
\aj, 102, 1381

\reference Villaver, E., Garc\'{\i}a-Segura, G., \& Manchado, A.\
2002a, \apj, 571, 880
% The dynamical evolution of the circumstellar gas ... I. The AGB

\reference Villaver, E., Garc\'{\i}a-Segura, G., \& Manchado, A.\
2003, ApJL, 585, 49
% Ram pressure stripping in PNe

\reference Villaver, E., Manchado, A., \& Garc\'{\i}a-Segura, G.\
2002b, \apj, 581, 1204
% The dynamical evolution of the circumstellar gas ... II. The PN formation


\end{references}
\end{document}